\documentclass[aps,groupedaddress] {revtex4-2}
\usepackage{amsmath,amssymb,amsfonts}
\usepackage{geometry}
\usepackage{graphicx}
\allowdisplaybreaks
\newcommand {\bp}{\begin{pmatrix}}
\newcommand {\ep}{\end{pmatrix}}
\newcommand{\be}{\begin{equation}} \newcommand{\ee}{\end{equation}}
\newcommand{\bea}{\begin{eqnarray}}\newcommand{\eea}{\end{eqnarray}}

\begin{document}

\title{
Construction of Pseudo-hermitian matrices describing systems with balanced loss-gain
}


\author{ Pijush K. Ghosh}
\email[]{pijushkanti.ghosh@visva-bharati.ac.in}
\affiliation{ Department of Physics, Siksha-Bhavana, Visva-Bharati University,
Santiniketan, PIN 731 235, India.}
\date{\today}

\begin{abstract}
We present a general construction of pseudo-hermitian matrices
in an arbitrary large, but finite dimensional vector space. The positive-definite
metric which ensures reality of the entire spectra of a pseudo-hermitian operator,
and is used for defining a modified inner-product in the associated vector space
is also presented. The construction for an $N$ dimensional vector space is based on the generators of $SU(N)$ in the
fundamental representation and the identity operator. We apply the results to
construct a generic pseudo-hermitian lattice model of size $N$ with balanced loss-gain.
The system is amenable to periodic as well as open boundary conditions and by construction,
admits entirely real spectra along with unitary time-evolution. The tight binding and
Su-Schrieffer-Heeger(SSH) models with nearest neighbour(NN) and next-nearest
neighbour(NNN) interaction with balanced loss-gain  appear as limiting cases.
\end{abstract}
\maketitle

\section{Introduction}

Pseudo-hermitian operators play an important role in the description of consistent quantum theory
for a class of non-hermitian Hamiltonians which admit entirely real spectra\cite{ali-review}. The advent
of ${\cal{PT}}$-symmetric quantum systems\cite{cmb} is followed by the developments in pseudo-hermitian operators,
and in a sense, these two concepts are complementary to each other. The basic idea of a pseudo-hermitian operator 
is to endow the vector space associated with a non-hermitian operator with a positive-definite metric such
that it becomes hermitian with respect to a modified inner product. The unitarity and completeness of states
are ensured with respect to the modified inner product. All non-hermitian operators are not amenable to such
a construction \textemdash the positive-definite metric can be found only for the operators satisfying the
condition of pseudo-hermiticity and admitting entirely real spectra. A prescription for constructing the
metric in the Hilbert space for a given non-hermitian Hamiltonian is well known which requires a priory
the complete knowledge of eigenspectra\cite{ali-review}. This requirement makes it difficult to find the
exact metric analytically for systems with infinite dimensional Hilbert space, and various approximate
methods have been used. The scenario for systems with finite-dimensional Hilbert space is better, and exact
metric has been constructed for a few specific physical models\cite{wang-1,wang-2,wang-3}.

The main focus of this article is the construction of generic pseudo-hermitian matrices and the associated metric
for arbitrary large, but finite dimensional vector space. The motivation stems from the fact that finite size
lattice models have representations in terms of matrices. The various types of tight binding models and spin chains
with non-hermitian interaction may be brought under the purview of such a construction. Pseudo-hermitian
matrices are also relevant for constructing a class of solvable models of vector non-linear Schr\"odinger equation,
non-linear Dirac equation and Oligomers with balanced loss-gain\cite{pkg-review, pkg-nlse}. Further, systems with
infinite
dimensional Hilbert space may be projected to specific sectors to obtain finite dimensional matrices \textemdash
multi-level quantum systems with non-hermitian interaction may be amenable for such studies. A construction
of pseudo-hermitian operators is also important in its own right from the viewpoint of mathematics.

The product of two hermitian matrices, with one of them being identified as the metric, satisfies the condition
of pseudo-hermiticity\cite{pkg-hami, piju-taming,yj,jf,jfmz}. A pseudo-anti-symmetric matrix $D=MR$ was used in the
Hamiltonian formulation of classical systems with balanced loss-gain, where the matrices $M, R, D$ correspond to
a generalized mass-term/metric, gauge-potential, loss-gain terms, respectively\cite{pkg-hami, piju-taming}.
The product-form
of the pseudo-hermitian matrix has been used in the study of pseudo-hermitian random matrix theory\cite{yj,jf}.
This product-form of pseudo-hermitian matrix is suitable for describing balanced loss-gain systems, since
the trace of the non-hermitian part of the matrix is identically zero.

In this article, we use the generators of $SU(N)$ and the $N \times N$ identity matrix as a basis to construct 
a hermitian matrix with $N^2$ independent real parameters. The product of two such hermitian matrices gives a
non-hermitian matrix depending on $2N^2$ independent real parameters. The resultant matrix is pseudo-hermitian
by construction, and we use the algebra of $SU(N)$ generators to express it in terms of
the basis formed by the $SU(N)$ generators and the identity matrix. A sufficient condition for the
metric to be positive definite is derived ensuring that the pseudo-hermitian matrix admits entirely
real spectra. The coefficients of the $SU(N)$ generators may be chosen judiciously to identify the
pseudo-hermitian matrix with some physical models. We construct a generic pseudo-hermitian lattice model
of size $N$ with balanced loss-gain. The system is amenable to both periodic as well open boundary conditions,
and admits entirely real spectra along with unitary time-evolution. We present a tight-binding model of size
$N$ with NN and NNN interaction. The lattice contains loss-gain terms at all sites and the strength of the
hopping term is non-uniform. As a special case, we obtain a tight binding Hamiltonian of size $N$ with
defects at the two end-points and uniform coupling. The positive-definite metric has been constructed. We
also present a SSH-type tight-binding model with explicit construction of the positive definite metric.

The plan of this article is the following. We present some basic results pertaining to pseudo-hermitian
operators\cite{ali-review} in the next section. The construction of the pseudo-hermitian operator is described in Sec.
II. A. A sufficient condition for the generic pseudo-hermitian operator to admit entirely spectra is also presented in this
section. Various known results related to the $SU(N)$ group are presented in order to facilitate  
model building from the generic pseudo-hermitian operator. In Sec. III, the generic pseudo-hermitian
matrix with NN and NNN interaction is presented. The balanced loss-gain terms are present at all
sites and hopping strengths are non-uniform. A uniform tight-binding model and a SSH-type model are
presented in Sec. III.A and III.B, respectively. Finally, results are summarized along with discussions
in Sec. IV.

\section{ Pseudo-hermitian Matrix}

We will be using pseudo-hermitian matrices in order to analyze a class of systems with balanced
loss and gain. We present the known basic results involving pseudo-hermitian matrices in order to
facilitate further discussions. An operator ${\cal{O}}$ is said to be pseudo-hermitian if it is
related to its adjoint through a similarity transformation,
\bea
{\cal{{O}}}^{\dagger} = \eta {\cal{O}} \eta^{-1}.
\eea
\noindent The hermitian operator $\eta$ is not unique for a given operator ${\cal{O}}$ and may or may not be
positive-definite. The operator ${\cal{O}}$ admits entirely real spectra and norm of the associated
eigen-vectors are non-negative if a  positive-definite $\eta:=\eta_+$ exists.
It should be noted that the operator ${\cal{O}}$ is non-hermitian in a vector-space with the standard
inner product. However, due to the pseudo-hermiticity property, it is hermitian in a vector space endowed
with a metric $\eta_+$ such that the scalar-product is defined as $\langle \langle \cdot , \cdot \rangle
\rangle_{\eta_+} \equiv \langle \cdot, \eta_+ \cdot \rangle$. The metric $\eta_+$ is hermitian
with respect to the standard inner-product and its positivity excludes
negative-norm states.  The standard results for a hermitian operator ${\cal{O}}$ is reproduced for
$\eta_+=I$, where $I$ is the identity operator.

A non-hermitian operator that is related to a hermitian operator through a similarity transformation is
known as quasi-hermitian operator\cite{scholtz}. The pseudo-hermiticity implies that ${\cal{O}}$ is also
crypto-hermitian, since the operator $\hat{\cal{O}}=\rho {\cal{O}} \rho^{-1}$ is hermitian with respect to
the standard inner-product in the associated vector-space. It should be noted that the operator
$\rho:=\sqrt{\eta_+}$ is well defined, since $\eta_+$ is positive-definite.
The defining relation for a pseudo-unitary operator\cite{ali} $V$ is $V^{\dagger} \eta_+ V=
\eta_+$ which can be realized in terms of a pseudo-hermitian operator ${\cal{O}}$ as $V=e^{-i {\cal{O}}}$.
The standard results for a unitary operator is reproduced for $\eta_+=I$ for which ${\cal{O}}$ is
hermitian. The identity $\rho^{-1} e^{- i \hat{\cal{O}}}= e^{-i {\cal{O}}} \rho^{-1}$ is  useful in various
contexts which implies that the action of the
operator $\rho^{-1}$ on the unitary operator $e^{- i \hat{\cal{O}}}$ is to change it to a pseudo-unitary
operator $e^{-i {\cal{O}}}$. Similarly, $\rho$ changes  $e^{-i {\cal{O}}}$ to $e^{- i \hat{\cal{O}}}$, i.e.
$\rho e^{-i {\cal{O}}}= e^{- i \hat{\cal{O}}} \rho$.

\subsection{Construction of Pseudo-hermitian Operator}

We consider three hermitian operators ${\cal{S}}, {{\eta}}$ and ${\cal{M}}$ in an $N$-dimensional
vector space with the condition $[\eta, {\cal{M}}]=0$ and $\eta$ is invertible. The pseudo-hermitian
operator ${\cal{O}}$ is constructed in terms of ${\cal{S}}, \eta$ and ${\cal{M}}$ as follows,
\bea
{\cal{O}} & = & a_0 {\cal{S \eta}} + a_1 {\cal{M}}\nonumber \\
& = & \underbrace{\frac{a_0}{2} \left \{{\cal{S}},{{\eta}} \right \}+a_1 {\cal{M}}}_{{\cal{O}}_h} +
\underbrace{\frac{a_0}{2} \left [ {\cal{S}},{{\eta}} \right ] }_{{\cal{O}}_{nh}},
\ \ a_0, a_1 \in \mathbb{R}\nonumber
\eea
\noindent where ${\cal{O}}_h$ is hermitian, while ${\cal{O}}_{nh}$ is non-hermitian. The matrix
${\cal{O}}$ is ${{\eta}}$-pseudo-hermitian, i.e. ${\cal{O}}^{\dagger}= {{\eta}} {\cal{O}} {{\eta}}^{-1}$, and
admits entirely real spectra for a positive-definite ${{\eta}}$.
The operator ${\cal{O}}_{nh}$ is trace less by construction, i.e. $Tr({\cal{O}}_{nh})=0$ and is
suitable for describing systems with balanced loss-gain terms. The operator ${\cal{O}}$ is defined
in terms of $3N^2+ 2$ real parameters \textemdash $2 N^2$ parameters corresponding to
hermitian matrices ${\cal{S}}$ and $\eta$, while $a_0, a_1, {\cal{M}}$ account for the extra $N^2+2$ parameters.
There are $N^2+2$ redundant parameters in this formalism and these parameters may be redefined
appropriately so that there are $2N^2$ independent real parameters as required for
a general pseudo-hermitian operator in $N$-dimensional vector space. The operator ${\cal{O}}$ is
pseudo-hermitian even for $a_1=0$. The sole purpose of adding the term is to have more flexibility
in constructing physical systems by using ${\cal{O}}$ with $a_1 \neq 0$. For example, it may be
convenient to define the metric $\eta$ in terms of total $p < N^2$ parameters to keep it simple.
The operator ${\cal{O}}$ with $a_1=0, a_0=1$ for such a choice is described in terms of $N^2 +p < 2N^2$
number of real parameters. The deficiency of $N^2-p$ parameters can be compensated partly or fully by
considering $a_1 \neq 0$. The operator ${\cal{M}}$ modifies the hermitian part ${\cal{O}}_h$ of ${\cal{O}}$ and it
may be chosen in various ways for a given $\eta$. A particular choice that is suitable for describing
a host of physical systems in terms of pseudo-hermitian operators is as follows:
\bea
{\cal{M}}= \sum_{k=1}^{N-1} b_k {{\eta}}^k, \ \  b_k \ \in \mathbb{R} \ \forall \ k.
\eea
\noindent The highest power of $\eta$ in ${\cal{M}}$ is $N-1$, since $\eta^k \ \forall \ k \geq N$ can
be expressed in terms of a polynomial in $\eta$ with the highest order $N-1$. This can be understood simply
in terms of matrix representation of the operator $\eta$ and Cayley-Hamilton theorem which states that
every matrix satisfies its own characteristic polynomial. We choose $a_0=1, a_1=0$ 
in rest of this article such that ${\cal{O}}$ is described in terms of $2N^2$ real parameters.
One may add an anti-hermitian operator $\xi$
to ${\cal{O}}$ with the condition $\{\xi, \eta \}=0$ for which ${\tilde{{\cal{O}}}={\cal{O}}+ \xi}$
is still $\eta$-pseudo-hermitian. However, the anti-commutation relation between $\eta$ and $\xi$
implies that $\eta$ is not positive-definite. Consequently,  ${\tilde{{\cal{O}}}}$ can not admit
entirely spectra and this possibility is abandoned.

The operators ${\cal{M}}$ and ${\eta}$ may be expressed as linear combinations
of $SU(N)$ generators $T^a, a=1, \dots, N^2-1$ and the identity operator $T^0$.
We use the normalization convention $Tr(T^a T^b)= 2 \delta^{ab}$, where $Tr(X)$
denotes the trace of $X$. The commutation and the ant-commutation relations among
$T^a$ are given by\cite{gm},
\bea
\left [ T^a, T^b \right ] = 2 i f^{abc} T^c, \ \ \left \{ T^a, T^b \right \}= \frac{4}{N} \delta^{ab}
+ 2 d^{abc} T^c,
\label{gen-identity}
\eea
\noindent where the completely anti-symmetric tensor $f^{abc}=\frac{1}{4i} Tr(\left [T^a,T^b\right] T^c)$
denote the structure constants and the completely symmetric tensor $d^{abc}=\frac{1}{4} Tr(\{T^a,T^b\}T^c)$. 
The operators $\eta$ and ${\cal{S}}$ are expressed in terms of $2N^2$ real parameters
$\alpha_a, \beta_a, a=0, 1, \dots, N^2-1$ as follows:
\bea
\eta & = & \sum_{a=0}^{N^2-1} \alpha_a T^a = \underbrace{\frac{1}{N} Tr(\eta)}_{\alpha_0} T^0 +
\sum_{a=1}^{n^2-1} \underbrace{\frac{1}{2} Tr(\eta T^a)}_{\alpha_a; a\geq1} \ T^a,\nonumber \\
{\cal{S}} & = & \sum_{a=0}^{N^2-1} \beta_a T^a =  \underbrace{\frac{1}{N} Tr({\cal{S}})}_{\beta_0} \ T^0 +
\sum_{a=1}^{n^2-1} \underbrace{\frac{1}{2} {Tr({\cal{S}} T^a)}}_{\beta_a;a\geq1} \ T^a.
\eea
\noindent The operator $\eta-\alpha_0 T^0$ is trace less and admits positive as well as negative eigenvalues.
The lowest negative eigenvalue of  $\eta-\alpha_0 T^0$ is denoted as $-\lambda_{min}, \lambda_{min} >0$ so that
the lowest eigenvalue of $\eta$ for $\alpha_0 >0$ is $\alpha_0-\lambda_{min}$. The condition
$\alpha_0 > \lambda_{min}$ ensures that $\eta$ is positive-definite. In absence of any closed form
expression for $\lambda_{min}$, a lower bound on $\alpha_0$ which ensures a positive-definite ${\eta}$
may be specified as,
\bea
\alpha_0 > \alpha_0^{min} \equiv \sqrt{\frac{1}{2} Tr((\eta-\alpha_0 T^0)^2)} =
\sqrt{ \sum_{a=1}^{N^2-1} (\alpha_a)^2 }
\eea  
\noindent The condition $\alpha_0 > \alpha_0^{min}$ is sufficient to ensure a positive-definite
$\eta$ and this is not a necessary condition. For all practical purposes, $\lambda_{min}$ may have
to be found analytically or numerically for a given $\eta$. The important message here is that it
is always possible to choose $\alpha_0$ such that ${\eta}$ is positive-definite. It is obvious from
the above discussions that $\eta$ can not be positive definite for $\alpha_0 \leq 0$. The expressions
for the operator ${\cal{O}}$ may be evaluated by using Eq. (\ref{gen-identity}) as,
\bea
{\cal{O}} & = & \sum_{a=0}^{N^2-1} \delta_a T^a\nonumber \\
& =& \underbrace{\left ( \alpha_0 \beta_0 + \frac{2}{N} \sum_{a=1}^{N^2-1} \alpha_a \beta_a \right )}_{\delta_0}
T^0 + \sum_{a=1}^{N^2-1} \underbrace{ \left ( \alpha_0 \beta_a + \beta_0 \alpha_a +
\sum_{b,c=1}^{N^2-1}  \left ( d^{abc}-i f^{abc} \right )
\alpha_b \beta_c \right )}_{\delta_a, a \geq 1} T^a.
\label{phm-gen}
\eea
\noindent The term $d^{abc}-i f^{abc}$ may be substituted as $\frac{1}{2} (Tr(T^aT^bT^c))^*$
by using the identity $Tr(T^aT^bT^c)=2 \left (d^{abc}+if^{abc}\right )$. The trace may be
computed analytically for a given representation of $SU(N)$ generators $T^a$ by using
any computer system algebra. The operator ${\cal{O}}$ is non-hermitian with
respect to the standard norm in the vector space due to the term containing the structure constants $f^{abc}$.
However, with modified inner-product in a vector space endowed with the metric $\eta$, ${\cal{O}}$
is a hermitian operator in terms $2N^2$ real parameters in an $N$ dimensional vector space. The associated
positive-definite metric operator $\eta$ with $N^2$ real parameters is also constructed. It may be noted that
previous attempts\cite{wang-1,wang-2,wang-3} in this regard in the context of ${\cal{PT}}$-symmetric and/or
pseudo-hermitian systems in $N$ dimensional Hilbert space produced non-hermitian Hamiltonian with total number
of real parameters less than $2N^2$. To the best of our knowledge, this is the first time that a construction
of the most general pseudo-hermitian operators with the maximum number of allowed real parameters is presented
for a given $N$. The large number of parameters allow more flexibility in constructing  physical systems in addition
to providing a complete formalism. Further, the known properties of $SU(N)$ generators may be useful for analyzing
the system. 

The explicit construction of pseudo-hermitian operator in $N$ dimensional vector space involves
choosing a representation of the $SU(N)$ generators $T^a$. We introduce a complete set of $N$
orthonormal basis vectors $|j\rangle$ satisfying the relations,
\bea
\langle j| k \rangle = \delta_{jk}, \ \ \sum_{j=1}^N |j\rangle \langle j| = I.
\label{su0}
\eea
\noindent The $N(N-1)/2$ symmetric generators $ \Lambda_{ij}^S$, $N(N-1)/2$ anti-symmetric
generators $\Lambda_{ij}^A$ and $N-1$ diagonal generators $\Lambda_{n^2-1}^D$ of $SU(N)$
in the fundamental representation can be expressed in terms of these basis vectors
as follows\cite{kimura,sunket}:
\bea
&& \Lambda^S_{k,j} = |k\rangle \langle j| + |j\rangle \langle k|, \ \
\Lambda^A_{k,j} = - i \left ( \ |k\rangle \langle j| - |j\rangle \langle k| \ \right ),
\ 1 \leq k < j \leq N,\nonumber \\
&& \Lambda^D_{n^2-1}= \sqrt{\frac{2}{n^2-n}} \left ( \sum_{j=1}^{n-1} |j\rangle\langle j|
- (n-1) |n\rangle\langle n| \right ), \ n=2, 3, \dots, N
\label{su1}
\eea
\noindent The $SU(N)$ generators are chosen to be orthogonal $Tr(\Lambda^a_{ij} \Lambda^b_{kl})= 2
\delta_{ab} \delta_{ik} \delta_{jl}$. The following identifications may be made between $T^a$ and
$\Lambda$ operators by expressing the index `$a$' of the generator $T^a$ in terms of $j, k$\cite{duncan}:
\bea
\Lambda^S_{kj}=T^{j^2+2(k-j)-1}, \ \Lambda^A_{kj}=T^{j^2+2(k-j)}, \ \ \Lambda^D_{n^2-1}=T^{n^2-1}
\ \textrm{for} \ n=2, \dots N
\label{su2}
\eea
\noindent The matrix representation of $\Lambda$ operators reproduce the generalized
Gell-Mann matrices\cite{gm}:
\bea
&& \left [ \Lambda_{k,j}^S \right ]_{lm} = \delta_{kl} \delta_{jm} + \delta_{km} \delta_{jl}, \ \
\left [ \Lambda_{k,j}^A \right ]_{lm} = -i \left ( \delta_{kl} \delta_{jm} -
\delta_{km} \delta_{jl} \right ), \ 1 \leq k < j \leq N \nonumber \\
&& \left [ \Lambda_{n^2-1}^D \right ]_{kl} = \sqrt{\frac{2}{n^2-n}} \bp I_{n-1,n-1} && {\bf O}_{n-1,N-n+1}\\
{\bf O}_{N-n+1,n-1} && P_{N-n+1,N-n+1} \ep,  \ n=2, \dots, N; \nonumber \\ 
&& \left [ P_{N-n+1,N-n+1} \right ]_{k^{\prime}l^{\prime}} = - (n-1) \delta_{k^{\prime}1}\delta_{l^{\prime}1},
\ k^{\prime}, l^{\prime}= 1,2, \dots, N-n+1, \ k,l=1, 2, \dots, N
\label{su3}
\eea
\noindent where $I_{m}$ and ${\bf O}_{mn}$ denote $m \times m$ identity and $m \times n$ null matrices,
respectively. Recently, closed form expressions for $f^{abc}, d^{abc}$ for any $N$ have been
given for the above representation of the $SU(N)$ generators\cite{duncan}, which may be used to evaluate
the term $d^{abc}-if^{abc}$ appearing in ${\cal{O}}$. We present our results in terms of $\Lambda^s,
\Lambda^a$ and $\Lambda^d_{n^2-1}$ and specific representations either in terms of ket-vectors or
matrices may be chosen depending on the physical scenario. 

\section{System with NN and NNN interaction}

The general construction of a pseudo-hermitian operator is
given in Sec. II. An analytical and/or numerical analysis of
a system with $2N^2$ parameters becomes non-trivial for large $N$.
There are plenty of physical systems which are described in terms
a few independent parameters. For example, the matrix representation
of ${\cal{O}}$ may correspond to a tight-binding model of size $N$ in which
each atom interacts with rest of the atoms on the lattice and defects are
present at all sites. The strength of the inter-atomic coupling decreases
with the increasing separation between two atoms. If the interaction among
the atoms is restricted to uniform NN and/or NNN coupling, there is a significant
reduction in the total number of independent parameters, yet it corresponds to a physical
model. Similar considerations for various types of lattice models of size $N$
or coupled discrete and continuum nonlinear Schr$\ddot{o}$inger equations are fruitful
in analyzing a large class of pseudo-hermitian systems in terms of a few real parameters.
In this section, we describe explicit construction of pseudo-hermitian operators which
may correspond to physical systems with NN and NNN interactions.

A general approach for constructing a physical model from Eq. (\ref{phm-gen}) would be to choose
the $2 N^2$ parameters appropriately such that it reduces to the desired form. A lattice model
which can be expressed in terms of a finite-dimensional matrix is amenable for this approach. For example, a
real symmetric tridiagonal matrix ${\cal{O}}-\delta_0 T^0$ with non-hermitian diagonal elements
is suitable for describing a tight binding Hamiltonian with balanced loss-gain and NN interaction.
This may be achieved  by imposing the following conditions:\\
\bea
&& \alpha_0 \beta_a + \beta_0 \alpha_a +
\sum_{b,c=1}^{N^2-1}  \left ( d^{abc}-i f^{abc} \right )
\alpha_b \beta_c = \delta_a \neq 0, \  a=k^2+2k-2 \ \textrm{and} \ a=n^2-1   \nonumber \\
&& \alpha_0 \beta_a + \beta_0 \alpha_a +
\sum_{b,c=1}^{N^2-1}  \left ( d^{abc}-i f^{abc} \right )
\alpha_b \beta_c = 0, \  a \neq k^2+2k-2 \ \textrm{and} \ a \neq n^2-1,
\label{condi}
\eea
\noindent where $ k=1, 2, \dots, N-1$ and $ n=2, 3, \dots, N$. It is apparent from
Eqs. (\ref{su0},\ref{su1},\ref{su2},\ref{su3}) that the first condition ensures non-vanishing
elements in the superdiagonal, subdiagonal and main diagonal, while the second condition implies that the
remaining elements are zero. All non-vanishing $\delta^a$'s may be considered to be equal for describing
a tight binding model with uniform coupling. Eq. (\ref{condi}) describes an underdetermined system \textemdash
there are $N^2-1$ equations in terms of $2 N^2$ variables. One may choose $\alpha_0, \beta_0, \beta_a$ as parameters
and $\alpha_a$'s as independent variables. This choice reduces Eq. (\ref{condi}) to a set of $N^2-1$ coupled linear
equations in terms of $N^2-1$ variables. Analytical solutions may be pursued for a few smaller values of
$N$. However, for large $N$, computer system algebra may be useful to find solutions. We present below
a tight binding model by following a different approach.

We consider metric $\eta$ and ${\cal{S}}$ as follows,
\bea
\eta  =  \gamma_0 I_N + \sum_{j=1}^{N} \gamma_j \Lambda^S_{j,j+1}, \ \
{\cal{S}}  =  \xi_0 I_N + \sum_{j=1}^{N} \xi_{j} \Lambda^A_{j,j+1}, \
\gamma_0, \xi_0, \gamma_i, \xi_i \ \in \mathbb{R} \ \forall i,
\eea
\noindent where the identifications $\Lambda^S_{N,N+j} =\Lambda^S_{j,N}, \
\Lambda^a_{N,N+j} =-\Lambda^a_{j,N}$ have been made. 
We introduce the complex parameters $\Gamma_j$ and the
operators $\Lambda_{k,j}^{+}, \Lambda_{k,j}^{-}$ as follows:
\bea
\Gamma_j=\xi_0 \gamma_j + i \xi_j \gamma_0, \
\Lambda_{k,j}^{+}=\frac{1}{2} \left ( \Lambda_{k,j}^S +
i \Lambda_{k,j}^A \right ) =|k\rangle\langle j|, \ \
\Lambda_{k,j}^{-}=(\Lambda_{k,j}^+)^{\dagger}=|j \rangle\langle k|, \ j > k.
\eea
\noindent The pseudo-hermitian operator ${\cal{O}}$ may be expressed as,
\bea
{\cal{O}} & = & \gamma_0 \xi_0 I_N + \sum_{j=1}^N \left ( \Gamma_j^* \Lambda_{j,j+1}^+
+ \Gamma_j \Lambda_{j,j+1}^- \right ) -i \sum_{j=1}^N \left ( \gamma_j \xi_{j+1}
\Lambda_{j,j+2}^+ - \gamma_{j+1} \xi_j \Lambda_{j,j+2}^- \right) \nonumber \\
& - & i \sum_{j=1}^N \left ( \gamma_j \xi_j - \gamma_{j+1} \xi_{j+1} \right )
\left [ \sqrt{\frac{j}{2(j+1)}} \Lambda^D_{(j+1)^2-1} -
\sum_{l=0}^{N-j-2} \frac{ \Lambda^D_{(N-l)^2-1}}{[2(j+l+1)(j+l+2)]^{\frac{1}{2}} } \right ]
\eea
\noindent where the identifications $\Lambda^D_{(N+j)^2-1}=\Lambda^D_{(j+1)^2-1}, \gamma_{N+j}=\gamma_j,
\xi_{N+j}=\xi_j$ have been made. The term with diagonal generators in ${\cal{O}}$ are non-hermitian
and correspond to balanced loss-gain. The second term of ${\cal{O}}$ is hermitian and corresponds to NN
interaction, while the third term describes non-hermitian NNN interaction for generic values of the parameters.
A particular physical scenario is that only balanced loss-gain terms are non-hermitian, while the remaining
terms of ${\cal{O}}$ are hermitian. This may be achieved by fixing the parameters as
$\xi_i = - C \gamma_i, C \in \mathbb{R},  i=1, \dots, N$ which make the terms describing NNN interaction
hermitian. The third term is still non-hermitian for this choice of parameters and describe loss-gain terms.
The complex parameters $\Gamma_i= \left ( \xi_0 - i C \gamma_0 \right ) \gamma_i$ for this particular
choice of $\xi_i$. The pseudo-hermitian operator ${\cal{O}}$ with balanced loss-gain is described in
terms $N+3$ real parameters as follows:
\bea
{\cal{O}} & = & \gamma_0 \xi_0 I_N + \sum_{j=1}^N \left ( \Gamma_j^* \Lambda_{j,j+1}^+
+ \Gamma_j \Lambda_{j,j+1}^- \right ) + i C \sum_{j=1}^N \gamma_j \gamma_{j+1} \left (
\Lambda_{j,j+2}^+ - \Lambda_{j,j+2}^- \right) \nonumber \\
& + & i C \sum_{j=1}^N \left ( \gamma_j^2 - \gamma_{j+1}^2 \right )
\left [ \sqrt{\frac{j}{2(j+1)}} \Lambda^D_{(j+1)^2-1} -
\sum_{l=0}^{N-j-2} \frac{ \Lambda^D_{(j+l+2)^2-1}}{[2(j+l+1)(j+l+2)]^{\frac{1}{2}} } \right ]\nonumber \\
& = & \gamma_0 \xi_0 + \sum_{j=1}^N \Big [ \Big ( \Gamma_j^* |j\rangle \langle j+1| 
+i C \gamma_j \gamma_{j+1} |j\rangle\langle j+2| + h.c. \Big )
- i C \Big ( \gamma_j^2 - \gamma_{j+1}^2 \Big ) |j+1\rangle\langle j+1| \Big ]
\label{ket-final}
\eea
\noindent where $|N+j\rangle=|j\rangle$ and h.c. denotes hermitian conjugate. The following
identity\cite{sunket} relating the matrix representations of the diagonal generators $\Lambda^D_{n^2-1}$
and the projection operator $|j\rangle\langle j|$ has been used in the last step of Eq. (\ref{ket-final}):
\bea
|j+1\rangle\langle j+1|= \frac{1}{N} I_N -\sqrt{\frac{j}{2(j+1)}} \Lambda^D_{(j+1)^2-1} +
\sum_{l=0}^{N-j-2} \frac{ \Lambda^D_{(j+l+2)^2-1}}{[2(j+l+1)(j+l+2)]^{\frac{1}{2}} },
j=0, \dots, N-1\nonumber
\eea 
\noindent The matrix representation of ${\cal{O}}$ is pentagonal sans the three corner elements
$[ {\cal{O}}_{h}]_{1,N-1}, [ {\cal{O}}_{h}]_{1,N}$ and $[ {\cal{O}}_{h} ]_{2,N}$.
The matrix is suitable for describing systems with nearest neighbour and next-nearest-neighbour interaction
with periodic boundary conditions. It may be noted that system with open boundary condition may also be
described by the same matrix by taking all three corner elements to be identically zero.
In particular, the choice $\gamma_N=\delta_N=0$ eliminates the corner elements and the system
is amenable for open boundary condition.

\subsection{Uniform coupling}

We present an example of a Pseudo-hermitian Tight-Binding Model with uniform couplings for NN and NNN terms
and two defects at the ends. We choose $\gamma_1= \gamma_2= \dots = \gamma_{N-1} \equiv \gamma, \gamma_N=0,
\xi_0=0$ for which ${\cal{H}} := \frac{1}{C\gamma_0} \left ( {\cal{O}}-\gamma_0 \delta_0 I_N \right )$
with ${\cal{O}}$ given in Eq. (\ref{ket-final}) takes the following form:
\bea
{\cal{H}} = i \gamma_0 \sum_{j=1}^{N-1} \Big [ |j\rangle \langle j+1 | - |j+1\rangle\langle j| \Big ]
+ i \gamma \sum_{j=1}^{N-2} \Big [ |j\rangle \langle j+2 | - |j+2\rangle\langle j| \Big ]
+ i \gamma \Big [ |1\rangle\langle 1| - |N\rangle\langle N| \Big ]
\eea
\noindent The first two terms of ${\cal{H}}$ are hermitian, while the last term is non-hermitian with
respect to the standard norm in the Hilbert space. The Hamiltonian is pseudo-hermitian with respect
to the metric $\eta$,
\bea
\eta=\gamma_0 \sum_{i=1}^N |i\rangle\langle i| + \gamma \sum_{i=1}^{N-1} \Big ( |i\rangle\langle i+1 | +
 |i+1\rangle\langle i | \Big )
\eea
\noindent which has the form of a tridiagonal matrix. In particular,  
$\eta =\gamma_0 I_N + \gamma \sum_{i=1}^{N-1} \Lambda_{i,i+1}^S$ and its eigenvalues are,
\bea
e_k=\gamma_0 + 2 \gamma \cos \left ( \frac{k \pi}{N+1} \right ), \ k=1, \dots, N.
\eea
\noindent Note that $e_i -\gamma_0 = -(e_{N+1-i} -\gamma_0)$ implying that all the roots are positive definite for
\bea
\gamma_0 > 2 {\vert \gamma \vert} \cos(\frac{\pi}{N+1})
\eea
\noindent which is the condition for a $\eta$-pseudo-hermitian ${\cal{H}}$ with entirely real spectra.
Denoting ${\cal{\eta}}_D= \textrm{diagonal}
(e_1, \dots, e_N)$ and $X^T$ as the transpose of $X$, the orthogonal matrix $O$ that diagonalizes ${\eta}$,
i.e. ${\eta}_D= O {\eta} O^T$, has the expression $[O]_{ij}=\sqrt{\frac{2}{N+1}} \sin(\frac{ij\pi}{N+1})$.
We define a matrix $\rho:=\sqrt{\eta}=O^T {{E}} O$, where $E=\textrm{diagonal} (\sqrt{e_1}, \dots, \sqrt{e_N})$.
The matrix $\rho$ can  be used to construct a hermitian Hamiltonian $h=\rho {\cal{H}} \rho^{-1}$ that is isospectral
with ${\cal{H}}$. We do not investigate the eigenspectra and physical properties of ${\cal{H}}$ in this article
which will be taken elsewhere\cite{SG-PG}.

\subsection{\bf SSH Model}

We present a pseudo-hermitian modified SSH model with defects at all sites. We choose $N=2m$,
$\gamma_{2i-1,2i}=\delta_1, \gamma_{2i,2i+1}=\delta_2, \xi_0=0$ for which the
Hamiltonian ${\cal{H}}:=({\cal{O}}-\gamma_0\xi_0 I_N)/C$ takes the form,
\bea
{\cal{H}} & = & i \gamma_0 \delta_1 \sum_{i=1}^m \Big ( |2i-1\rangle\langle 2i| 
- |2i\rangle\langle 2i+1| \Big ) + i \gamma_0 \delta_2 \sum_{i=1}^m \Big ( |2i\rangle\langle 2i+1| -
 |2i+1\rangle\langle 2i| \Big )\nonumber \\
&  + & i \delta_1 \delta_2 \sum_{i=1}^N \Big ( |i\rangle \langle i+2| - |i\rangle \langle i+2| \Big )
 + i \Big ( \delta_2^2 -\delta_1^2 \Big ) \sum_{i=1}^N (-1)^{i+1} |i\rangle\langle i|
\eea
\noindent The Hamiltonian contains NNN interaction in addition to the balanced loss-gain. The Hamiltonian
is pseudo-hermitian with respect to the metric $\eta$,
\bea
\eta=\gamma_0 \sum_{i=1}^N |i\rangle \langle i| + \delta_1 \sum_{i=1}^m \Big ( |2i-1\rangle \langle 2i|
+ |2i\rangle \langle 2i-1| \Big )
+\delta_2 \sum_{i=1}^{m} \Big ( |2i\rangle\langle 2i+1| + |2i+1\rangle\langle 2i| \Big )
\eea
\noindent The condition $\gamma_0 > \sqrt{m(\delta_1^2 + \delta_2^2)}$ is sufficient to ensure that
the eigenvalues of $\eta$ are positive-definite. 

\section{Conclusions \& Discussions}

We have presented a construction of a generic $N \times N$ pseudo-hermitian matrix in terms of
$2N^2$ real parameters. This is the most general form of an $N \times N$ pseudo-hermitian matrix.
The explicit realization of such a matrix utilizes the generators of $SU(N)$ group, the
identity matrix $I_N$ and their properties. We have also presented the explicit form of the
metric in the associated vector space which is required to define a modified norm. A sufficient
condition has been stated ensuring a positive-definite metric, and hence entirely real spectra
for the pseudo-hermitian matrix. 

The generic pseudo-hermitian matrix is suitable for describing various types of lattice models
appearing in physics. We have presented a generic pseudo-hermitian tight-binding model of size
$N$ with NN and NNN interaction, and defects at all sites leading to balanced loss-gain. The
strength of the hopping term is non-uniform for this model. Further, we have specialized to
uniform coupling and defects at the two extreme end-points. The metric for the tight-binding
model and the condition for its positivity have also been presented. We have also presented
a SSH-type model as a special case of the generic model.

There are a few new directions related to the present construction. For example, the
eigen-spectra, transport properties, topological properties, $\dots$, etc. of the tight
binding models presented in this article have not been investigated in this article. It
will be reported in a separate publication\cite{SG-PG}. Further,
it may be interesting to derive other physically interesting lattice models from the generic
pseudo-hermitian matrix. It can be applied to the construction of exactly solvable vector
non-linear Schr\"odinger equation, non-linear Dirac equation, Oligomer etc. by following
the methods outlined in Ref. \cite{pkg-review}. Some of these issues will be addressed
in future publications.

\end{document}